\documentclass[aps,twocolumn,showpacs,superscriptaddress]{revtex4-1}  
\usepackage{graphicx}  
\usepackage{dcolumn}   
\usepackage{bm}        
\usepackage{amssymb}   
\usepackage{color}

\newcommand{\blue}{\color{blue}}
\begin{document}

\widetext
\leftline{arXiv: accepted by {\blue Phys. Rev. Lett.}}

\title{\blue Explosive Turbulent Magnetic Reconnection}

\author{K. Higashimori}
\affiliation{Department of Earth and Planetary Science, University of Tokyo}
\author{N. Yokoi}
\affiliation{Institute of Industrial Science, University of Tokyo}
\author{M. Hoshino}
\affiliation{Department of Earth and Planetary Science, University of Tokyo}
\date{\today}

 \begin{abstract}
  We report simulation results for turbulent magnetic reconnection
  obtained using a newly developed Reynolds-averaged
  magnetohydrodynamics model.
  We find that the initial Harris current sheet develops in three
  ways, depending on the strength of turbulence:
  laminar reconnection, turbulent reconnection, and turbulent
  diffusion.
  The turbulent reconnection explosively converts the magnetic field
  energy into both kinetic and thermal energy of plasmas, and generates
  open fast reconnection jets.
  This fast turbulent reconnection is achieved by the localization of
  turbulent diffusion.
  Additionally, localized structure forms through the 
  interaction of the mean field and turbulence.
 \end{abstract}

\maketitle


  Critical questions relating to magnetic reconnection are
  how and when does fast reconnection take place,
  particularly in the case of
  a high magnetic Reynolds number ($R_m\sim 10^{10}$). 
  Because a magnetic Reynolds number is sufficiently high to maintain
  turbulence in space, much attention has been paid to the
  relationship between turbulence and magnetic reconnection {\blue \cite{Matthaeus2011}}. 
  For example, it has been theoretically suggested that in strong
  Alfv${\rm \acute e}$nic turbulence, the reconnection rate does not
  depend on electric resistivity, but rather on the properties of
  turbulence (such as the characteristic scale length and power of the
  fluctuation) {\blue \cite{Lazarian1999}}.
  The scenario has been examined in magnetohydrodynamics
  (MHD) simulations, where external turbulence is forced in a current
  sheet {\blue \cite{Kowal2009}}.
  The effect of turbulence on the reconnection rate has also been
  investigated in large-scale two-dimensional simulation;
  it was suggested that there is a critical turbulent power above which
  turbulence greatly affects reconnection and the reconnection
  rate has weaker dependence on electric resistivity than
  does Sweet--Parker {\blue \cite{Sweet1958,Parker1957}} reconnection
  {\blue \cite{Loureiro2009}}.
  Another study focused on reconnection in turbulence in terms of a
  turbulent cascade, and it was suggested that reconnection rates are
  distributed and controlled by turbulence {\blue \cite{Servidio2009}}.
  Recently, another viewpoint on the relationship between turbulence and
  reconnection has been presented according to the theory of MHD turbulence
  {\blue \cite{Yokoi2011a}}.
  In that study, it was theoretically suggested that the generation of
  cross-helicity ${\cal W}\equiv \left<\bm{v}'\cdot\bm{b}'\right>$
  (where $\bm{v}'$ and $\bm{b}'$ are respectively the characteristic
  velocity and magnetic field of the turbulent motion) dramatically
  enhance the rate of reconnection.

  We apply a Reynolds-averaged MHD model and
  investigate the nonlinear evolution of turbulent reconnection.
  In the model, physical quantities (such as the velocity $\bm{v}$) are
  decomposed into mean and turbulent quantities; e.g.,
  $\bm{v}=\bm{V}+\bm{v'}$ (where the capital letter stands for the mean
  quantity).
  In this study, tubulent effects are included in Ohm's law.
  Taking the ensemble average, $\left<...\right>$ (e.g.,
  $\left<\bm{v}\right>=\bm{V}$), of the Ohm's law
  gives a following equation:
  \begin{eqnarray}
   \bm{E}=\eta \bm{J}-\bm{V}\times \bm{B}
   -\left<\bm{v}'\times \bm{b}'\right>,\ \label{eq5}
  \end{eqnarray}
  where $\left<\bm{v}'\times \bm{b}'\right>$ is the electromotive force
  due to effects of turbulence, and it modulates the mean electric
  field.
  Mean variables, such as $\bm{V}$ and $\bm{B}$, are obtained using
  ordinal MHD equations:
  \begin{eqnarray}
   \frac{\partial \rho}{\partial t}
    +\bm{\nabla}\cdot\left(\rho \bm{V}\right)=0\label{eq1},\\
   \frac{\partial}{\partial t}\left(\rho \bm{V}\right)
   +\bm{\nabla}\cdot\left[\rho \bm{V}\bm{V}+\left(P+\frac{B^2}{2}\right)
		{\bf I}-\bm{B}\bm{B}\right]=\bm{0}\label{eq2},\\
   \frac{\partial}{\partial t}
   \left(\frac{P}{\gamma_{a}-1}
    +\frac{\rho}{2}V^2+\frac{B^2}{2}
	    \right)\hspace*{7mm}\nonumber\\
   +\bm{\nabla}\cdot\left[
		\left(\frac{\gamma_{a}}{\gamma_{a}-1}P
		 +\frac{\rho}{2}V^2\right)\bm{V}
		+\bm{E}\times \bm{B}
	       \right]=0,\label{eq3}\\
   \frac{\partial \bm{B}}{\partial t}+\bm{\nabla}\times
   \bm{E}=\bm{0},\label{eq4}
  \end{eqnarray}
  where $\bf{I}$ is the unit tensor.
  Note that the velocity is normalized by the Alfv$\acute{\rm e}$n velocity.
  $\gamma_{a}$ is an adiabatic index and set to $\gamma_{a}=5/3$.

  The most important part of modeling turbulence is the evaluation of
  the electromotive force, $\left<\bm{v}'\times \bm{b}'\right>$, which
  reflects the information of turbulence.
  Here, we applied the model for inhomogeneous MHD turbulence
  {\blue \cite{Yoshizawa1990,Yoshizawa1993,Yokoi2013b}}.
  In this model, the electromotive force can be written as
  \begin{eqnarray}
   \left<\bm{v}'\times \bm{b}'\right>=
    -\beta_{t} \bm{J} + \gamma_{t} \bm{\Omega},\label{eq6}
  \end{eqnarray}
  where $\bm{\Omega}=\bm{\nabla}\times \bm{V}$ is the mean vorticity.
  $\beta_{t}$ and $\gamma_{t}$ are respectively expressed as
  $\beta_{t}= C_\beta\tau {\cal K}$ and $\gamma_{t}= C_\gamma\tau {\cal W}$.
  ${\cal K}\equiv\left<v'^2+b'^2\right>/2$ and
  ${\cal W}= \left<\bm{v}'\cdot\bm{b}'\right>$ are respectively the
  macroscopically defined turbulent energy and cross-helicity.
  $\tau$ is the characteristic time scale of turbulence.
  $C_\beta$ and $C_\gamma$ are model constants of $O(10^{-1})$.
  (We checked that the result did not strongly depend on their exact
  values.)
  We set $C_\beta=C_\gamma=0.3$ in the present paper.
  The advantage of this model is that the coefficients $\beta_{t}$ and
  $\gamma_{t}$ are determined by the following equations for turbulence
  {\blue \cite{Yokoi2008}}:
  \begin{eqnarray}
   \frac{\partial{\cal K}}{\partial t}=
    -\left<\bm{v}'\times \bm{b}'\right>\cdot \bm{J}
   +\bm{B}\cdot\bm{\nabla}{\cal W}
   -\bm{V}\cdot\bm{\nabla}{\cal K}
   -\epsilon_{\cal K}
   \label{eq7},\\
   \frac{\partial{\cal W}}{\partial t}=
   -\left<\bm{v}'\times \bm{b}'\right>\cdot \bm{\Omega}
   +\bm{B}\cdot\bm{\nabla}{\cal K}
   -\bm{V}\cdot\bm{\nabla}{\cal W}
   -\epsilon_{\cal W}
   \label{eq8},
  \end{eqnarray}
  where $\epsilon_{\cal K}$ and $\epsilon_{\cal W}$ are respectively the
  dissipation rates of turbulence energy and cross-helicity.
  They are given by
  \begin{eqnarray}
   &\epsilon_{\cal K}=\displaystyle \frac{\cal K}{\tau}\label{eq9},\\
   &\epsilon_{\cal W}=\displaystyle C_{\cal W}\frac{\cal W}{\tau},\label{eq10}
  \end{eqnarray}
  where $C_{\cal W}$ is a model constant {\blue \cite{Yokoi2011b}}, and set to
  $1.3$ here.
  The above set of equations (\ref{eq5})--(\ref{eq10}) is numerically
  solved using the two-dimensional ($x$--$z$ plane) fourth-order Runge--Kutta
  and fourth-order central difference scheme {\blue \cite{Rempel2009}}.
  Grid intervals $\Delta x$ and $\Delta z$ are fixed to unity, and
  the simulation size is $L_x\times L_z=2048\times 512$.
  (The simulation box ranges $\left|x\right|/L_z\leq 2.0$ and
  $-0.25\leq z/L_z\leq0.75$.)
  Boundaries in $x$- and $z$-directions are both periodic, and a pair of
  Harris current sheets is assumed.
  Hereafter, only the region $\left|z\right|/L_z\leq 0.25$ (i.e., the lower
  current sheet) is discussed.

  In this simulation model, we must give the initial conditions for both
  the mean field (such as $\bm{B}$ and $P$) and the turbulent
  field (such as ${\cal K}$ and ${\cal W}$).
  As for the initial mean field, the mean magnetic field for the lower
  current sheet ($\left|z\right|/L_z\leq 0.25$) is given by
  $\bm{B}=B_{x0}\tanh\left(z/\delta\right)\bm{e}_x 
  +B_{z0}\sum_{m=1}^{10}\sin\left(2\pi mx/L_x\right)\bm{e}_z$,
  where $B_{x0}=1.0$ and $B_{z0}/B_{x0}=1.0\times 10^{-3}$.
  $\delta (=0.02 L_z)$ is half-thickness of the initial current sheets.
  $\bm{e}_x$ and $\bm{e}_z$ are respectively the unit vectors in $x$-
  and $z$-directions.
  The plasma beta outside the current sheets is set to 
  $\beta_{p}=0.5$, and the spatial distribution of $P$ is determined to
  satisfy the pressure balance.
  Additionally, uniform electric resistivity, $\eta=1.0\times
  10^{-2}$, is assumed to avoid numerical instability.

  As for the initial turbulent field, we assume ${\cal W}=0$
  and ${\cal K}=1.0\times 10^{-2}$ everywhere.
  We found that the magnitude of ${\cal K}$ did not change the basic
  property of turbulent reconnection.
  In addition, a steady state of turbulent energy,
  $\partial {\cal K}/\partial t=0$,
  is assumed in the initial current sheets.
  Because $\bm{\Omega}=\bm{V}=\bm{0}$ and ${\cal W}=0$ at the center of
  the current sheets, it holds that
  $\tau_0=C_\beta^{-1/2}\left|\bm{J}\right|^{-1}_{z=0}$ in the steady
  state.
  (Here equations (\ref{eq6}), (\ref{eq7}), and
  (\ref{eq9}) are used.)
  To investigate the relationship between turbulence and
  reconnection, we slide $\tau$ from the steady state,
  $\tau=\tau_0$, by introducing the parameter $C_\tau$ as
  $\tau=C_\tau \tau_0$.
  It should be noted that $\tau$ is constant throughout each
  simulation run.
  We execute simulations with different $C_\tau$ values, and the
  characteristic cases A--D are summarized in Table \ref{tab:1}.
  (The total number of simulation runs corresponds to the number of red
  points in Figure \ref{fig2}.)
  Since $C_\tau$ determines the characteristic time scale of
  turbulence, it controls dissipation rate of
  turbulent energy, $\epsilon_{\cal K}$.
  For example, if $C_\tau$ is much smaller than unity,
  $\epsilon_{\cal K}$ becomes $\epsilon_{\cal K}\gg 1$, and
  ${\cal K}\rightarrow 0$ is expected as the simulation time passes.
  In such a case, the system will develop into laminar flow.
  On the other hand, in the case of $C_{\tau}\gg 1$, it is expected that
  the system quickly becomes turbulent, and reconnection will not
  occur owing to strong turbulent diffusion.
  (Here it is noted from equations (\ref{eq5}), (\ref{eq4}), and
  (\ref{eq6}) that the $\beta_t$-related term in the induction equation is
  $\partial \bm{B}/\partial t=\cdots-\bm{\nabla}\times\left[\left(\eta+\beta_t\right)\bm{J}\right]$,
  and the turbulent diffusion of the mean magnetic field increases as
  $\cal K$ increases.)
  Therefore, in this study, we focus on the most interesting parameters around
  $C_\tau\sim 1$.
  \begin{table}
   \caption{\label{tab:1}Simulation parameters}
   \begin{ruledtabular}
    \begin{tabular}{cccccccc}
     Run& A &$\cdots$&B&$\cdots$&C&$\cdots$ &D\\
     \hline
     $C_\tau$&0.05&$\cdots$& 0.5&$\cdots$&1.2&$\cdots$ &3.0\\
    \end{tabular}
   \end{ruledtabular}
  \end{table}

  \begin{figure}
   \includegraphics[scale=0.3]{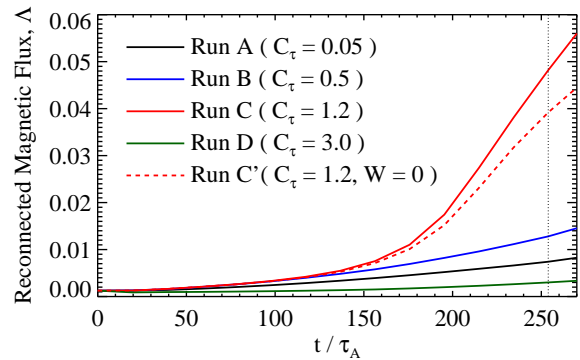}
   \caption{\label{fig1} Time evolution of the reconnected magnetic
   fluxes in the cases of $C_\tau=0.05$ (black),
   $C_\tau=0.5$ (blue), $C_\tau=1.2$ (red), and
   $C_\tau=3.0$ (dark green).
   In Run (C') (red-dashed), $\cal W$ is forced to be $0$ throughout the
   calculation.
   }
  \end{figure}

  \begin{figure}
   \includegraphics[scale=0.3]{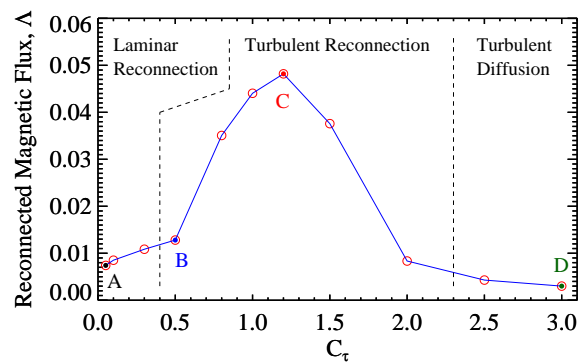}
   \caption{\label{fig2} Reconnected magnetic fluxes at $t/\tau_A=254$
   vs $C_\tau$.
   Red points stand for simulation runs with different $C_\tau$ values.
   }
  \end{figure}

  \begin{figure*}
   \includegraphics[scale=0.3]{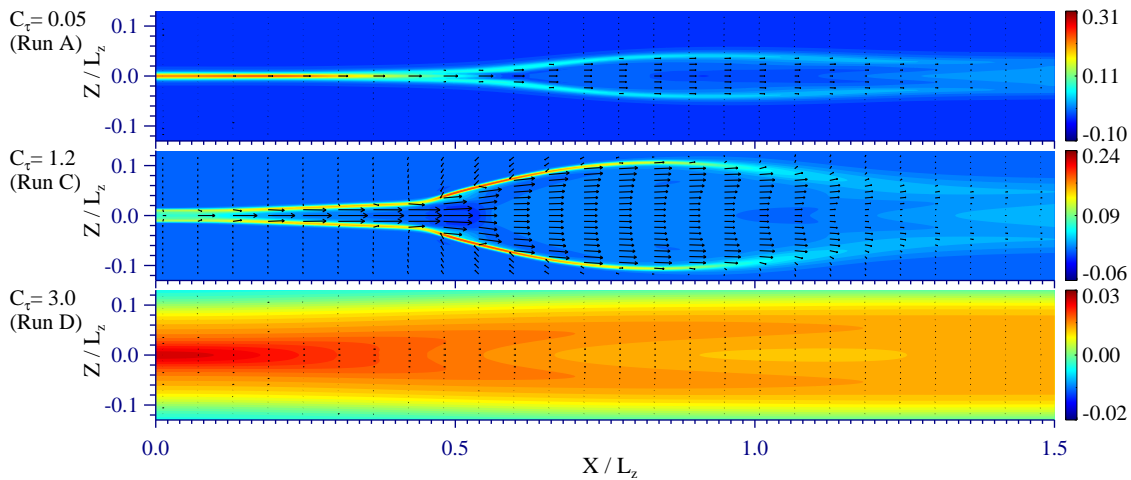}
   \caption{\label{fig3} $y$ components of the electric currents in
   the laminar ($C_\tau=0.05$) and turbulent ($C_\tau=1.2,\ 3.0$)
   cases at $t/\tau_A=254$ are shown as color contour plots.
   Black arrows show the flow velocity.
   }
  \end{figure*}

  Figure \ref{fig1} shows the time evolution of the reconnected magnetic
  flux,
  $\Lambda\equiv\int_{-L_x/2}^{+L_x/2}dx\left|B_z\right|_{z=0}/(B_{x0}L_x)$,
  for simulation runs (A)--(D).
  (In Run (C'), the cross helicity, $\cal W$, is switched off in order to
  discuss the contribution of $\cal W$ to the reconnection rate. This is
  referred to later.)
  Time is normalized by the Alfv${\rm \acute e}$n transit time,
  $\tau_A\equiv\delta /V_A$.
  In the present paper, we do not discuss the evolution at time
  $t/\tau_A>260$ to avoid the boundary effect on the reconnection
  {\blue \cite{Forbes1987}}.
  The black line shows the reconnected flux, $\Lambda$, in the
  case of $C_\tau=0.05$.
  In this case, both $\beta_t$ and $\gamma_t$ soon drop to zero, and the
  reconnection becomes slow laminar one.
  As $C_\tau$ increases, reconnection develops more quickly, and
  the development is fastest
  around $C_\tau\sim 1.2$ (see the red line) for the present simulation
  parameters.
  However, with larger $C_\tau$, the speed of reconnection again drops (see
  the dark-green line), and falls below that of laminar
  reconnection.
  In this case, the turbulent energy $\cal K$ quickly increases, and the
  resultant strong turbulent diffusion in the initial current sheet
  prevents reconnection.
  As a result, no outflow jet is observed. (The profile of the
  velocity is shown in the bottom panel of Figure \ref{fig3} as black
  arrows.)
  Figure \ref{fig2} shows the relationship between reconnected magnetic
  fluxes and the parameter $C_\tau$ at $t/\tau_A=254$ for all simulation
  runs.
  The figure clearly shows that the initial current
  sheet develops in three ways depending on $C_\tau$ values and the
  speed of reconnection is fastest for a moderate $C_\tau$ value
  ($C_\tau\sim 1$).

  We now discuss the spatial structures of the above three cases.
  In Figure \ref{fig3}, the $y$ components of the electric current
  at $t/\tau_A=254$ in the laminar ($C_\tau=0.05$) and turbulent
  ($C_\tau=1.2,\ 3.0$) cases are shown as contour plots.
  Flow vectors are overlaid as black arrows.
  In the laminar case, a Sweet--Parker-type current sheet forms
  {\blue \cite{Biskamp1986,Uzdensky2000}}, and reconnection is more gradual than
  in the turbulent case.
  Under strong turbulence ($C_\tau=3.0$), the initial
  current sheet (with thickness $\sim 0.04\ z/L_z$) quickly broadens and
  reconnection does not take place, as shown in the bottom panel.
  On the other hand, in the case of turbulent reconnection ($C_\tau=1.2$: see
  the middle panel), two pairs of current sheets 
  formed as Petschek-type reconnection {\blue \cite{Petschek1964}}, and open
  reconnection jets are observed.
  The existence of Petschek-type reconnection itself has already
  supported by the results of anomalous resistivity models
  {\blue \cite{Sato1979,Ugai1992}}, however, the inherent physical processes of
  turbulent reconnection are quite different from them.
  It should manifest that the physics is self-consistently
  determined by the nature of MHD turbulence.

  \begin{figure}
   \includegraphics[scale=0.3]{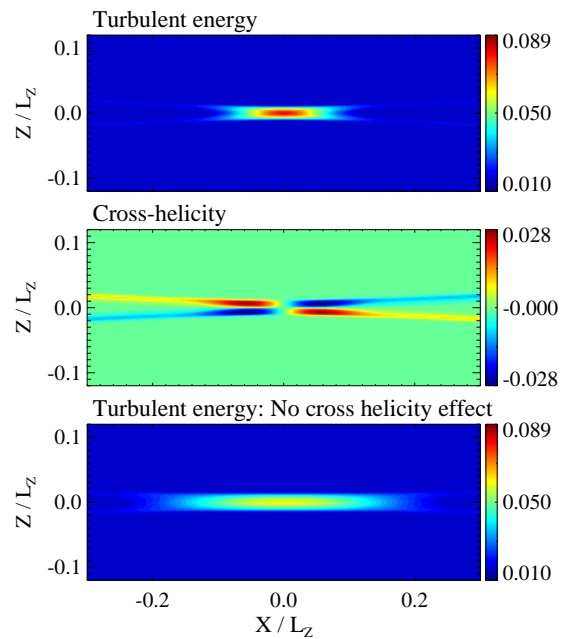}
   \caption{\label{fig4}
   The upper two contour plots show the spatial
   distributions of the turbulent energy ${\cal K}$ and the
   cross-helicity ${\cal W}$ near the magnetic neutral point
   in the case of $C_\tau=1.2$.
   The bottom panel shows the spatial distribution of $\cal K$ at
   $C_\tau=1.2$, where $\cal W$ is switched off throughout the
   calculation.
   These three snapshots are taken at time $t/\tau_A=254$.
   }
  \end{figure}

  In such fast reconnection cases, both the turbulent energy $\cal K$
  and cross-helicity
  $\cal W$ are efficiently produced, and the turbulent diffusion near
  the magnetic neutral point is locally strengthened.
  The upper two panels of Figure \ref{fig4} show the spatial
  distributions of the turbulent energy $\cal K$ and cross-helicity
  $\cal W$ near the turbulent diffusion region at time
  $t/\tau_A=254$ in the case of $C_\tau=1.2$.
  These spatial distributions of $\cal K$ and $\cal W$ are
  determined by the interaction between the mean and turbulent fields.
  In the first stage of turbulent reconnection, as $\bm{V}\sim\bm{0}$ and 
  ${\cal W}\sim 0$, the production term of equation (\ref{eq7}),
  $-\left<\bm{v}'\times\bm{b}'\right>\cdot\bm{J}\sim \beta_t \bm{J}^2$,
  dominates.
  The increasing $\beta_t=C_\beta\tau{\cal K}$ then facilitates turbulent
  diffusion and triggers reconnection.
  Subsequently, topological change in the magnetic field and vorticity
  develops.
  The electric current and vorticity generate the cross-helicity,
  according to the production term of equation (\ref{eq8});
  i.e., $-\left<\bm{v}'\times\bm{b}'\right>\cdot\bm{\Omega}
  =\beta_t\bm{J}\cdot\bm{\Omega}-\gamma_t\bm{\Omega}^2$.
  In this way, a quadrupole structure of $\cal W$ forms.
  Then, $\cal W$ plays an important role on the development of
  the localized turbulent diffusion region.
  In the absence of $\cal W$, the turbulent diffusion region broadens as
  is shown in the bottom panel of Figure \ref{fig4}.
  On the other hand in the presence of $\cal W$, $\cal K$ is locally
  strengthened and the reconnection rate increases compared to that in
  the ${\cal W}=0$ case (see red-solid and red-dashed lines in Figure
  \ref{fig1}).

  In this paper, the relation between turbulence and
  magnetic reconnection was investigated using
  the Reynolds-averaged MHD model, where mean and turbulent
  fields develop by interacting with each other.
  It was found that the initial current sheet develops in three ways:
  laminar reconnection, turbulent reconnection, and turbulent diffusion.
  Reconnection develops most quickly in the second case, and it macroscopically
  appears as a single X-type reconnection with open fast outflow jets
  owing to the locally strengthened turbulent diffusion.
  Such fast turbulent reconnection would, for example, play an
  important role in the region where there is a large gap between the
  overall and dissipation scales.
  It is not clear whether reconnection around the dissipation
  scale really develops into a huge-scale phenomenon.
  Assuming the existence of turbulence, however, the thickness of the
  current sheet does not necessarily become as thin as the dissipation
  scale, if $\beta_t>\eta$.
  Instead of dissipation-scale physics, turbulent diffusion
  could macroscopically change the topology of (mean) magnetic fields.
  On the other hand, in the case that the scale gap is comparatively
  small, kinetic effects should be taken into account.
  From the viewpoint of turbulent reconnection, the outflow of
  collisionless magnetic reconnection could become turbulent
  without any forced perturbation {\blue \cite{Daughton2011}}.
  Additionally, along the boundary between inflow and outflow regions,
  Alfv${\rm \acute{e}}$n waves could be driven by the ion beam
  accelerated around the diffusion region {\blue \cite{Higashimori2012}}.
  We suggest that such self-generated turbulence and waves would
  macroscopically appear as sources of $\cal K$ and $\cal W$, and may
  further enhance the turbulent diffusion and the resultant reconnection
  rate.

  The model allows us to investigate phenomena of MHD
  turbulence even in the case of a high magnetic Reynolds number,
  and we hope that this will contribute to studies where direct
  numerical simulation (DNS) is difficult or impossible.
  On the other hand, it should be mentioned that the model provides less
  accuracy than DNS.
  For example, in the present paper, the characteristic turbulent
  timescale $\tau$ is assumed to be constant, and this may result in
  overestimation of turbulent diffusion in the case of $C_\tau\gg 1$.
  The timescale of turbulence, as well as the dissipation rate of
  turbulent energy, should be determined according to the nonlinear
  dynamics of turbulence.
  In future works, the accuracy of the model needs to be improved
  through mutual understanding with DNS and observations to
  clarify the role of turbulence in various phenomena.

  This research is supported by Japan Society for the Promotion of
  Science (JSPS; Grant No. 12J10000, 22001, and 24540228), and partially
  supported by NAOJ and NORDITA.

%


\end{document}